\documentstyle[preprint,prb,eqsecnum,aps]{revtex}

\begin{document}
\tightenlines
\draft
\author{Boldizs\'ar Jank\'o, Anders Smith$^*$, and Vinay Ambegaokar}
\address{Laboratory of Atomic and Solid State Physics, Cornell University,
Ithaca, New York 14850-2501, USA}
\date{\today}
\title{BCS Superconductivity with Fixed Number Parity}
\maketitle
\begin{abstract}
We investigate  superconductivity
in a grand canonical ensemble with
{\it fixed number parity} (even or odd). In the low temperature limit we
find small corrections to the BCS gap equation and dispersion $E(k)$.
The even-odd free energy difference in the same limit decreases linearly with
temperature, in accordance with the behavior observed experimentally
and previously arrived at from a quasiparticle model. The
theory yields deviations from the BCS predictions for the specific heat,
ultrasound attenuation, NSR relaxation rate, and electromagnetic absorption.
\end{abstract}

\pacs{PACS numbers: 74.20.Fg, 05.30.Fk}

\section{Introduction}
\label{sec:intro}

An extremely interesting recent development is the experimental
study of capacitatively isolated superconducting metallic
islands to which electrons can be added one by one.  Two different
experimental probes of such islands\cite{tuominen,devoret} have
demonstrated that even
when the number of electrons is extremely large---of the order of
$10^9$---a periodicity is observed for every two electrons added.
Since the microscopic BCS theory of superconductivity\cite{schrieffer}
is a pairing theory,
it is natural to assume that the observed alternation is due to
the evenness or oddness of the number of electrons on the island.
At first sight one might have guessed that adding electrons would involve
a localized state near the injection site, first filling it and
then emptying it to form a pair.  However, the
temperature and magnetic field dependence of the effect confirm that
for the reported experiments the relevant energies relate to
superconducting correlations.  Since these experiments probe
fundamental properties of superconductors their precise analysis is
important.  What makes this problem particularly interesting from the
point of view of theory is the fact that superconducting order is
off-diagonal in number so that the treatment of fixed number is
problematical.

Tuominen, Hergenrother, Tighe, and Tinkham \cite{tuominen} have suggested
that equilibrium properties can be calculated from the partition functions
\begin{equation}
{\cal Z}_{e/o}={1\over 2}\left\{\prod_{k,\sigma}
\big[1+e^{- \beta E_{k,\sigma}}\big]
\pm \prod_{k,\sigma}\big[1-e^{- \beta E_{k,\sigma}}\big]\right\},\label{5}
\end{equation}
where $e$ means even and $o$ odd, and $E_{k,\sigma}$ are the quasiparticle
energies.  The sign $\pm$ retains only configurations containing even or
odd numbers of quasiparticles in a grand canonical ensemble.  This
approach has also been adopted by Lafarge et al. \cite{devoret} in
interpreting their experimental results.
Now, Eq.(\ref{5}) cannot be quite right because the BCS theory is a
self-consistent theory in which the energy is
not simply a sum of
quasiparticle energies.  On the other hand, the number parity of a
state with $N$ quasiparticles is the same as that of a state with $N$
electrons.  In this paper we explore the
consequences of treating the islands by considering the appropriate
generalization of (\ref{5}), namely
\begin{equation}
{\cal Z}_{e/o}={1\over 2} {\rm Tr }\left\{\big[1\pm (-1)^N\big]
e^{-\beta({\cal H}-\mu N)}\right\}.
\label{6}
\end{equation}
We obtain the free energy for both
even and odd case and show that in the low temperature limit it reduces to the
ansatz (\ref{5}).  Using the same framework, we calculate
the ultrasound attenuation, NSR relaxation rate, and electromagnetic
absorption and find corrections to the standard BCS results at
low temperature.

The method of  Eq.(\ref{5}) is normally used in another context
where fixed number is important: nuclear
physics \cite{fetterwal}. There energy-levels are calculated using an
unrestricted grand canonical ensemble but only those with the  appropriate
quantum number for the nucleus in question are kept.  Since we are here
interested in thermodynamic properties which are  related to
fluctuations, it would seem more correct at the outset to suppress states
with the wrong number parity in the statistical ensemble, as in (\ref{6}).

In the BCS theory one fixes the average number of electrons
 $N$  by adjusting the chemical potential. The
fluctuations in particle number (of order $\sqrt{(N\Delta / E_F)}$ at
$T=0$, where $\Delta$ is the energy gap and $E_F$ the Fermi energy)
are negligible for a macroscopic system.  Since we are interested in a
small free energy difference in the present work, we keep
track of number fluctuations in both the even and the odd ensembles
of (\ref{6}), and explicitly check that they do not affect the
physical properties we calculate.

This paper is organized as follows: in Sec.(\ref{sec:free}) we calculate
the partition functions for the parity restricted ensembles; in
Sec.(\ref{sec:min}) we minimize the grand potential and obtain the gap
equation and dispersion for even or odd parity. We find only
small corrections to the BCS results.
Sec.(\ref{sec:thermo}) is devoted to extracting thermodynamic properties
from the grand potential. The differences between the results
obtained with the method of Eq.(\ref{6}) and the nuclear physics
method, Eq.(\ref{5}), turn out too small to be measurable with the
experimental techniques presently available. All subsequent calculations
are made within the approximation (\ref{5}). We evaluate
the even/odd free energy difference paying special attention to number
fluctuations. The entropy and specific heat of the even and odd parity
systems are also calculated: the results show deviations from
BCS.
In Sec.(\ref{sec:trans})
we show how the number parity restriction affects the response
of an even-  or odd-N superconductor to various external probes. The
final Sec.(\ref{sec:concl}) contains a summary and discussion.

\section{Calculation of Partition Function}
\label{sec:free}

In principle the most suitable framework for discussing the problem would be to
consider N particles interacting via the BCS interaction and construct a
BCS-like trial ground-state wave function and excited states with a
{\it fixed } number of particles \cite{irkin}.  For our purposes,
however, this scheme runs into technical difficulties.
The fixed-N BCS wave function cannot be normalized exactly, and a saddle
point approximation yields the grand-canonical BCS results.

If one tries \cite{smith} to
use the even and odd ground states in the usual grand-canonical method, the
difference between them will be practically removed by the parity-violating
grand-canonical process of a single-particle exchange between the system and
the particle reservoir.

In the partition function (\ref{6}) the projection operator
$[1 \pm (-1)^N/ 2]$ suppresses states with odd and even number parity,
respectively. The expectation value of an
operator in such an ensemble has the form
\begin{equation}
\langle O \rangle _{e/o} =
{\cal Z}_{e/o} ^{-1}{\rm Tr } \lbrace \frac{1 \pm (-1)^N}{2} O
e^{- \beta ({\cal H} - \mu N)} \rbrace . \label{av}
\end{equation}
The reduced BCS  Hamiltonian in the expressions above,
\begin{equation}
{\cal H} - \mu N = \sum _{k, \sigma } \epsilon _k c_{k \sigma }^{\dagger } c_{k
\sigma }
	- \sum _{k k' } V_{k k' } c_{k \uparrow }^{\dagger }
	c_{ -k \downarrow }^{\dagger }c_{ -k' \downarrow }
        c_{ k' \uparrow } ,
\end{equation}
(where the single-particle energies are measured from $\mu$)
can be transformed, using the Bogoliubov-Valatin transformation
\cite{rick}
\begin{eqnarray}
\gamma _{k0} = u_k c_{ k \uparrow } -  v_k c_{ -k \downarrow }^{\dagger }, \\
\gamma _{k1} = v_k c_{ k \uparrow }^{\dagger } + u_k c_{ -k \downarrow }, \\
u_k^2 + v_k ^2 = 1,
\end{eqnarray}
into the expression
\begin{equation}
{\cal H} - \mu N = {\cal H}_Q + {\cal H}_I + U .
\label{htot}
\end{equation}
Here ${\cal H}_Q$ corresponds to the kinetic energy of the Bogoliubov
quasiparticles
\begin{equation}
{\cal H}_Q = \sum _k E_k ( \gamma _{k0}^{\dagger } \gamma _{k0} +
		    \gamma _{k1}^{\dagger } \gamma _{k1} ) .
\label{quasih}
\end{equation}
where $E_k $ will be obtained later on by minimizing the grand potential.
Note that if (\ref{quasih}) were substituted into (\ref{6}), Eq. (\ref{5})
would result. However, there are two additional terms in (\ref{htot}).
${\cal H}_I$ describes the interaction between the elementary excitations.
\begin{eqnarray}
{\cal H}_I = \sum _k \left[ \epsilon_k ( u_k^2 - v_k ^2)
        + 2 u_k v_k \sum _{k'} u_{k'} v_{k'} V_{k k' }
        -  E_k \right] ( \gamma _{k0}^{\dagger } \gamma _{k0} +
                    \gamma _{k1}^{\dagger } \gamma _{k1} )  \nonumber \\
\FL     - \sum _{k, k'} V_{k k'} u_k v_k u_{k'} v_{k'}
        ( \gamma _{k0}^{\dagger } \gamma _{k0} +
        \gamma _{k1}^{\dagger } \gamma _{k1} )
         ( \gamma _{k'0}^{\dagger } \gamma _{k'0} +
         \gamma _{k'1}^{\dagger } \gamma _{k'1} ).
\end{eqnarray}
Finally, $ U$  is a c-number term depending on temperature
:
\begin{equation}
U = 2 \sum _k \epsilon _k v_k^2 - \sum_{k,k'} V_{k k'}  u_k v_k u_{k'} v_{k'}.
\end{equation}
We now make the BCS mean field approximation by assuming that the
low temperature dynamic properties of the system are governed by the
independent quasiparticles described by ${\cal H}_Q$ , and that the
interaction
term can be replaced by its thermal average
\begin{equation}
{\cal H} - \mu N \simeq {\cal H}_Q + C_{e/o},
\end{equation}
where the temperature dependent constant $C_{e/o}$ is  defined as
\begin{eqnarray}
\FL
C_{e/o} = \langle {\cal H}_I \rangle_{e/o} + U  = \nonumber \\
	= \frac{{\rm Tr } \lbrace \frac{1 \pm (-1)^N}{2} {\cal H}_I
e^{- \beta {\cal H}_Q} \rbrace}{{\rm Tr } \lbrace \frac{1 \pm (-1)^N}{2}
e^{- \beta {\cal H}_Q} \rbrace }
+ 2 \sum _k \epsilon _k v_k^2 - \sum_{k,k'} V_{k k'}  u_k v_k u_{k'} v_{k'}.
\end{eqnarray}
In this approximation it is possible to calculate the even-odd partition
function explicitly. The result can be given in terms of the
quantities
\begin{eqnarray}
{\cal Z}_{\pm } = \prod _{k, \sigma} ( 1 \pm e^{- \beta E_k} ) \label{zpm},\\
f_{\pm }(E_k) = \frac{ \pm 1}{e^{ \beta E_k} \pm 1} \label{fpm}.
\end{eqnarray}
Here $f_+(E_k)$ is the usual Fermi function. Then the even-odd partition
function has the form
\begin{equation}
{\cal Z}_{e/o} = \frac{1}{2} ({\cal Z}_+ \pm {\cal Z}_-) e^{ - \beta C_{e/o}} .
\label{part}
\end{equation}
In the same mean field approximation $C_{e/o}$ becomes
\begin{eqnarray}
C_{e/o} = C_{+} + \delta C_{e/o}, \nonumber \\
\delta C_{e/o} = \pm \frac{{\cal Z}_- ( C_{-}
- C_+)}{{\cal Z}_+ \pm {\cal Z}_-}.
\label{deltac}
\end{eqnarray}
The function $ C_+ $ in the above expression is the BCS result for
the constant C,
\begin{eqnarray}
C_+ = 2 \sum_k \epsilon _k v_k^2 +
2 \sum_k \left[ \epsilon _k (u_k^2 - v_k^2) - E_k \right] f_+(E_k)
\nonumber \\
- \sum_{k,k'} V_{k k'}  u_k v_k u_{k'} v_{k'} \left[ 1 - f_+(E_k) \right]
\left[ 1 - f_+(E_{k'}) \right],
\end{eqnarray}
whereas $C_- $ can be obtained from $C_+$ by a replacement of
$f_+(E_k)$ with $f_-(E_k)$,
\begin{equation}
C_- = C_+ [f_+(E_k) \rightarrow f_-(E_k)].
\end{equation}
Inserting these relations into Eq.(\ref{deltac}) we find
\begin{eqnarray}
\FL \delta C_{e/o} =
- \frac{2}{1 \pm \prod _{k, \sigma } \coth \frac{ \beta E_k}{2} } \Big\lbrace
\sum _k \frac{ \left[ \epsilon _k (u_k^2 - v_k^2 )- E_k \right]}
{ \sinh \beta E_k}
+ 2 \sum _{k, k'} V_{k k'}  u_k v_k u_{k'} v_{k'} \frac{ \coth  \beta E_k }
{\sinh \beta E_{k'} }\Big\rbrace .\nonumber \\
\end{eqnarray}
The even-odd grand potential can be directly obtained from the
even-odd partition function via ${\Omega }_{e/o} \equiv
- k_B T \ln {\cal Z}_{e/o}$. Casting ${\Omega }_{e/o}$ in a form that
makes the even-odd corrections explicit, we finally obtain
\begin{eqnarray}
{\Omega }_{e/o} = {\Omega }_{BCS} - \frac{1}{\beta } \ln \bigg[ \frac{1}{2}
\Big( 1 \pm \prod _{k, \sigma } \tanh \frac{ \beta E_k}{2} \Big) \bigg]
+ \delta C_{e/o}.
\nonumber \\
\label{f}
\end{eqnarray}

The last two terms in Eq. (\ref{f})
represent corrections arising from even-odd effects. These corrections
vanish for large enough temperature and/or volume. The easiest way to see how
this happens when the temperature is increased is to calculate the product
\begin{equation}
\prod _{k, \sigma} \tanh \frac{ \beta E_k}{2}  \equiv e^{ \sum _{k, \sigma}
\ln \tanh \frac{ \beta E_k}{2}} .
\end{equation}
For $ \Delta \ll k_B T $, so that $ E_k \simeq |\epsilon _k |$, one  obtains
\begin{equation}
\prod _{k, \sigma} \tanh \frac{ \beta E_k}{2}  = e^{- {3 \pi ^2 N}/
{2 \beta \epsilon _F}}.
\end{equation}
where $N$ is the total number of particles and $ \epsilon _F $ is the
Fermi energy. Since all even-odd corrections are proportional to
some power of this product, such effects
are  negligible at temperatures close to $T_c$.
As we will see, a much lower temperature
limit for the onset of even-odd effects can be set if we start from the low
temperature limit and calculate the temperature where the even-odd free
energy difference vanishes. In fact, when $ \Delta \gg k_B T $,
the relevant excitations are $ E_k \simeq \Delta $.
Defining $N_{eff}$ by \cite{tuominen}
\begin{equation}
\prod _{k, \sigma} \tanh \frac{ \beta E_k}{2}  \equiv
\bigg( \tanh \frac{ \beta \Delta }{2} \bigg) ^{N_{eff}},
\end{equation}
direct calculation shows that
\begin{eqnarray}
\FL N_{eff} \simeq 4 N(0) {\cal V } \int _{\Delta } ^{\infty } dE
\frac{E}{\sqrt{E^2 - \Delta ^2 }} e^{ - \beta (E - \Delta )}
\simeq 2 N(0) {\cal V} \sqrt{2 \pi k_B T \Delta } ,
\label{neff}
\end{eqnarray}
where ${\cal V}$ is the volume of the system.
Since $\tanh x < 1$ for all finite arguments $x$, the product is
again negligible for large enough $N_{eff}$, i.e. large volume or
particle number. For the samples used by Tuominen et al. and
Lafarge et al.  $N_{eff} \sim 10^4$.

\section{Minimization of the  grand potential}
\label{sec:min}

Before performing the minimization it is useful to rearrange
the terms in Eq.(\ref{f}) as follows
\begin{equation}
\Omega _{e/o} = \Omega _{BCS} (f_+(E_k) \rightarrow f(E_k)_{e/o}) -
4 \sum_{k,k'} V_{k k'}  u_k v_k u_{k'} v_{k'}
\left[ \langle n_k n_{k'} \rangle _{e/o} -
\langle n_k \rangle _{e/o} \langle n_{k'} \rangle _{e/o} \right],
\label{fmin}
\end{equation}
where $n_k = \gamma _k ^{\dagger } \gamma _k $  is the quasiparticle
number operator and
\begin{equation}
 f(E_k)_{e/o} \equiv \langle n_k \rangle _{e/o} =
\frac{f_+(E_k) {\cal Z}_+ \pm f_-(E_k) {\cal Z}_-}{{\cal Z}_+ \pm {\cal Z}_-}.
\end{equation}
Also
\begin{eqnarray}
\FL \Omega _{BCS} (f_+(E_k) \rightarrow f(E_k)_{e/o}) =
- \frac{1}{\beta } \ln \bigg[ \frac{1}{2}
\Big( {\cal Z}_+ \pm {\cal Z}_- \Big) \bigg]
+ \sum_{k, \sigma} \epsilon _k v_k^2
\nonumber \\+  \sum_{k, \sigma}
\left[ \epsilon _k (u_k^2 - v_k^2) - E_k \right] f_{e/o}(E_k)
- \sum_{k,k'} V_{k k'}  u_k v_k u_{k'} v_{k'}
\left[ 1 - 2 f_ {e/o}(E_k) \right] \left[ 1 - 2 f_{e/o} (E_{k'}) \right].
\end{eqnarray}
As we shall see later on, the last term in Eq.(\ref{fmin}) is small, but
nonzero. This is an explicit manifestation of the fact that the thermal Wick's
theorem is weakly violated in the even/odd ensembles.
The grand potential is a functional of two independent functions, say $ E_k$
and $v_k$, ${\Omega}_{e/o} = {\Omega}_{e/o}( \{ E_k, v_k\} )$.
Minimization with respect to these parameters will give two variational
equations, which in the BCS case yield the energy spectrum and the gap
equation.
Following  the same strategy, we obtain
\begin{eqnarray}
2 \epsilon_k u_k v_k - (u_k^2 - v_k^2) \Delta _k^{e/o} =
A_k^{e/o} \label{gap},\\
E_k - \epsilon _k (u_k^2 - v_k^2) - 2 u_k v_k \Delta _k^{e/o} = B_k^{e/o}
\label{disp}.
\end{eqnarray}
Here $ \Delta _k^{e/o} $ is defined as
\begin{eqnarray}
\Delta _k^{e/o} \equiv \sum _{k'} V_{k,k'} u_{k'} v_{k'}
[1 - 2 f_{e/o}(E_{k'})].
\label{gapeq}
\end{eqnarray}
Note that $u_k$ and $v_k$ are also parity dependent.
We have omitted the e/o labels for the sake of simplicity.
We shall see later on that the gap in the spectrum is somewhat
different from $ \Delta _k^{e/o} $.

Both $A_k^{e/o}$ and $B_k^{e/o}$  were generated by the
residual, Wick's theorem violating term in Eq.(\ref{fmin}).
\begin{eqnarray}
\FL
A_k^{e/o} = 4 \frac{u_k^2 - v_k^2}{1 - 2 f_ {e/o}(E_k)}
\sum_{k,k'} V_{k k'}  u_k v_k u_{k'} v_{k'}
\left[ \langle n_k n_{k'} \rangle _{e/o} -
\langle n_k \rangle _{e/o} \langle n_{k'} \rangle _{e/o} \right],
\label{a}
\end {eqnarray}
\begin{eqnarray}
\FL
B_k^{e/o} = 4 \frac{u_k v_k}{ -  \partial f_ {e/o}(E_k)/ \partial E_k}
\sum_{k,k'} V_{k k'}  u_k v_k u_{k'} v_{k'} \frac{\partial }{\partial E_k}
\left[ \langle n_k n_{k'} \rangle _{e/o} -
\langle n_k \rangle _{e/o} \langle n_{k'} \rangle _{e/o} \right].
\label{b}
\end {eqnarray}
These coefficients also vanish in the thermodynamic BCS limit, since then,
for reasons similar to those given above for the grand potential, even/odd
effects are small, and Wick's theorem is restored. Also note that, since
$A_k^{e/o}$ and $B_k^{e/o}$ are proportional to $u_k^2 - v_k^2$ and
$u_k v_k$, respectively, they can be interpreted as changes in
$ \Delta _k^{e/o} $ due to Wick's theorem violation.

\subsection{Exact solution of the minimization equations}
\label{sec:ex}

It is possible to give a formal exact solution to (\ref{gap}) and
(\ref{disp}) by making the usual change of variables $u_k = \cos \theta _k ,
v_k = \sin \theta _k $.
We obtain from the first equation of minimization, (\ref{gap}) a solution
similar to the BCS case
\begin{eqnarray}
u_k = \sqrt{\frac{1}{2}\biggl(1 + \frac{\epsilon _k}
{ (\epsilon _k^2 + {\overline \Delta  ^2 _{e/o}} )^{1/2}}\biggr)}, \\
v_k = \sqrt{\frac{1}{2}\biggl(1 - \frac{\epsilon _k}
{ (\epsilon _k^2 + {\overline \Delta  ^2 _{e/o}})^{1/2}}\biggr)},
\end{eqnarray}
but now ${\overline \Delta _{e/o}}$ is given by
\begin{eqnarray}
{\overline \Delta _k^{e/o}} =  \sum _{k'} V_{k,k'}
\frac{{\overline \Delta _k^{e/o}}}
{2 (\epsilon _k^2 + {\overline \Delta  ^2 _{e/o}})^{1/2}}\big\lbrace
[1 - 2 f_{e/o}(E_{k'})] \nonumber\\
+ \frac{1}{1 - 2 f_ {e/o}(E_k)}
\left[ \langle n_k n_{k'} \rangle _{e/o} -
\langle n_k \rangle _{e/o} \langle n_{k'} \rangle _{e/o} \right] \big\rbrace,
\label{gapeq2}
\end{eqnarray}
The spectrum $E_k$ can be obtained from (\ref{disp}) by using the above
results.
\begin{eqnarray}
E_k = (\epsilon _k^2 + {\overline \Delta ^2 _{e/o}})^{1/2} -
\frac{ {\overline \Delta _{e/o}}}{(\epsilon _k^2 +
{\overline \Delta ^2 _{e/o}})^{1/2}}
\sum _{k'} V_{k,k'}
\frac{{\overline \Delta _k^{e/o}}}
{2 (\epsilon _k^2 + {\overline \Delta  ^2 _{e/o}})^{1/2}} \nonumber \\
\times
\Bigl(\frac{2}{1 - 2 f_ {e/o}(E_k)}+
\frac{1}{  \partial f_ {e/o}(E_k)/ \partial E_k}
\frac{\partial }{\partial E_k} \Bigr)
\left[ \langle n_k n_{k'} \rangle _{e/o} -
\langle n_k \rangle _{e/o} \langle n_{k'} \rangle _{e/o} \right]
\label{disp2}
\end{eqnarray}

The expressions above contain the quantity
\begin{eqnarray}
\rho_{e/o} (k, k') \equiv \langle n_k n_{k'} \rangle _{e/o} -
\langle n_k \rangle _{e/o} \langle n_{k'} \rangle _{e/o}.
\end{eqnarray}
 This can be evaluated explicitly. For $k \ne k'$
\begin{eqnarray}
\rho_{e/o} (k, k') = \frac{f_+(E_k) f_+(E_{k'}){\cal Z}_+
\pm f_-(E_k) f_-(E_{k'}){\cal Z}_-}
{{\cal Z}_+ \pm {\cal Z}_-}  - \nonumber \\
\Biggl(\frac{f_+(E_k) {\cal Z}_+ \pm f_-(E_k) {\cal Z}_-}
{{\cal Z}_+ \pm {\cal Z}_-} \Biggr)
\Biggl(\frac{f_+(E_{k'}) {\cal Z}_+ \pm f_-(E_{k'}) {\cal Z}_-}
{{\cal Z}_+ \pm {\cal Z}_-}\Biggr).
\end{eqnarray}
Inserting the expressions for $f_{\pm}$ and ${\cal Z}_{\pm}$
[ Eq.(\ref{fpm} - \ref{zpm}) ], we get
\begin{eqnarray}
\rho_{e/o} (k, k') = \frac{{\rm cosech} \beta E_k \ {\rm cosech} \beta E_{k'}}
{\bigl(1 \pm \prod _{k, \sigma } \coth \frac{ \beta E_k}{2}\bigr)
\bigl(1 \pm \prod _{k, \sigma } \tanh \frac{ \beta E_k}{2}\bigr)}.
\label{rho}
\end{eqnarray}
On the other hand, for $k=k'$  $\rho (k, k) $ has the usual  form for
noninteracting fermions
\begin{eqnarray}
\rho_{e/o} (k, k') = f_{e/o}(E_k) - f_{e/o}(E_k)^2
\end{eqnarray}
The results obtained so far are exact. The
expressions for ${\overline \Delta _k^{e/o}}$ [ Eq. (\ref{gapeq2})] and the
dispersion $E_k$ [ Eq. (\ref{disp2})] constitute a closed system of two
self-consistent equations which in principle can be solved iteratively.
It is, however, possible to give approximate analytic result for
the low temperature limit.

\subsection{Low temperature approximation of the exact solution}
\label{sec:aprox}

The temperature range where the experiments\cite{tuominen,devoret}
revealed even/odd effects is $ 50  {\rm mK } < T < T^*
\simeq 200 {\rm mK}$. In this temperature range, for the given island
parameters $N_{eff} \sim 10^4 $. Then the following inequalities hold:
\begin{equation}
\exp ( - \beta \Delta ) \ll \frac{1}{N_{eff}} \ll 1 .
\end{equation}
Here $\Delta $ is the smallest value of the dispersion $E_k$ for both the
even and odd cases. As we shall see the zero temperature limit of the
gap in the even and the odd case differs, but by  a small amount only.
So in first approximation $\Delta $ is the same for both parities.
These inequalities enable us to systematically keep track of the terms in
an expansion of the exact result and to make a controlled approximation.
In this regime
\begin{eqnarray}
\frac{1}{1 \pm \prod_{k, \sigma} \coth \frac{ \beta E_k}{2}} \simeq
\frac{1}{1 \pm ( 1 + 2 N_{eff} e^{ - \beta {\overline \Delta ^0 _{e/o}}})}, \\
\frac{1}{1 \pm \prod_{k, \sigma} \tanh \frac{ \beta E_k}{2}} \simeq
\frac{1}{1 \pm ( 1 - 2 N_{eff} e^{ - \beta {\overline \Delta ^0 _{e/o}}})}.
\end{eqnarray}
Here ${\overline \Delta ^0 _{e/o}} \equiv {\overline \Delta  _{e/o}}(T=0)$
The parity dependent average occupation number $f_{e/o}(E_k)$  then
becomes
\begin{eqnarray}
f_{e/o}(E_k) = f_+(E_k) - \frac{ {\rm cosech } \beta E_k}
{1 \pm \prod _{k \sigma} \coth \frac{\beta E_k}{2}} \simeq \nonumber \\
\exp ( - \beta E_k) - \frac{ 2 \exp ( - \beta E_k) }
{1 \pm ( 1 + 2 N_{eff} e^{ - \beta {\overline \Delta ^0 _{e/o}}})}.
\end{eqnarray}
For the even parity we have
\begin{equation}
f_{e}(E_k) = N_{eff} \exp [ - \beta (E_k + {\overline \Delta ^0 _{e}})],
\end{equation}
whereas for the odd case
\begin{equation}
f_{o}(E_k) =  \exp (- \beta E_k )\Bigl( 1  +
\frac{ \exp \beta {\overline \Delta ^0 _{e}}}{N_{eff}} \Bigr).
\end{equation}
Similar calculation leads to approximate results for $\rho_{e/o} (k, k')$.
We obtain for $k \ne k'$
\begin{eqnarray}
\rho _e (k, k') = \exp [ -\beta (E_k + E_{k'}) ] ,
\sim {\cal O} \exp (- 2\beta \Delta ) \\
\rho _o (k, k') = \frac{\exp [ -\beta (E_k + E_{k'} - 2
{\overline \Delta ^0 _{o}}) ]}{N_{eff} ^2} \sim {\cal O} (1/N_{eff}^2).
\end{eqnarray}
We see that $\rho_{e/o} (k, k') $ is smaller than the even/odd corrections
coming from $f_{e/o}(E_k)$ by at least a factor of $1/N_{eff}$, and
therefore can be safely neglected in Eq. (\ref{gapeq2}) and
Eq. (\ref{disp2}).
The contribution from the case $k = k'$ is also negligible, since
it represents only one term in a summation over $k'$. Furthermore
\begin{equation}
\frac{1}{ -  \partial f_ {e/o}(E_k)/ \partial E_k}
\frac{\partial \rho _{e/o} (k, k')}{\partial E_k} =
\pm \frac{\exp [ -\beta ( E_{k'}  - {\overline \Delta ^0 _{e/o}})]}{N_{eff}}.
\end{equation}
Finally, using the results we obtained above, it is possible
to give a somewhat simpler, approximate expression for
${\overline \Delta  _{e/o}}$ and $E_k$ .
\begin{eqnarray}
{\overline \Delta _k^{e/o}} =  \sum _{k'} V_{k,k'}
\frac{{\overline \Delta _k^{e/o}}}
{2 (\epsilon _k^2 + {\overline \Delta  ^2 _{e/o}})^{1/2}}
[1 - 2 f_{e/o}(E_{k'})] + {\cal O} \bigl(\frac{1}{N_{eff}^2} \bigr).
\label{gapeq3}
\end{eqnarray}
\begin{eqnarray}
\FL
E_k = (\epsilon _k^2 + {\overline \Delta ^2 _{e/o}})^{1/2} \nonumber \\
\mp \frac{ {\overline \Delta ^2 _{e/o}}}
{(\epsilon _k^2 + {\overline \Delta ^2 _{e/o}})^{1/2}}
\sum _{k'}
\frac{V_{k,k'} \exp [ -\beta ( E_{k'}  - {\overline \Delta ^0 _{e/o}})]}
{2 N_{eff} (\epsilon _k^2 + {\overline \Delta  ^2 _{e/o}})^{1/2}}
+ {\cal O} \bigl(\frac{1}{N_{eff}^2} \bigr).
\label{disp3}
\end{eqnarray}
Notice that within this approximation ${\overline \Delta _k^{e/o}} =
\Delta _k^{e/o}$. From Eq.(\ref{gapeq3}) we can now extract\cite{landau}
the low temperature behavior of $ \Delta  _{e/o} $. Using
$E_k \simeq (\epsilon _k^2 +  \Delta ^2 _{e/o})^{1/2}$, we obtain for the
even case
\begin{equation}
\Delta_{e}(T) \simeq \Delta_{BCS} (0) \Bigl[ 1 -
\bigl(4\pi k_B T  N(0) {\cal V} \bigr)
\exp (- 2 \beta \Delta_{e} (0)) \Bigr].
\end{equation}
Here $\Delta_{BCS} (0)$ is the zero temperature limit of
the BCS gap for the same interaction $V_{k,k'} $.
Note that $\Delta_e (0) = \Delta_{BCS} (0)$. We used the expression
(\ref{neff}) for $N_{eff}$ to obtain the prefator of the exponential.
The most significant difference between
the above expression and the usual BCS result is the
$\exp (- 2 \beta \Delta )$ decrease of the gap, to be contrasted
with the $\exp (-  \beta \Delta )$ behavior predicted by the conventional
BCS theory. Furthermore the prefactor of the exponential is linear in $T$,
whereas in the BCS is proportional to $T^{1/2}$. For odd parity
\begin{equation}
\Delta_{o}(T) \simeq \Delta_{BCS}(0) \Bigl[ 1
- \frac{1}{2 \Delta_{o}(0)  N(0) {\cal V}}
- \bigl(\frac{2\pi k_B T}{ \Delta_{o} (0)}\bigr)^{1/2}
\exp (-  \beta \Delta_{o} (0)) \Bigr].
\end{equation}
Note that in the odd case $\Delta_{o}(T)$ saturates to a value that is
less than the BCS gap:
\begin{equation}
\Delta_{o}(0) \simeq \Delta_{BCS}(0)
- \frac{1}{2   N(0) {\cal V}  }.
\end{equation}
The difference, however, is very small, $ \sim 10^{-4}\Delta _{BCS }$.
Finally, let us examine Eq.(\ref{disp3}) which gives the approximate
dispersion. We can evaluate the sum in the second term and obtain
\begin{equation}
\sum _{k'}
\frac{V_{k,k'} \exp [ -\beta ( E_{k'}  - \Delta ^0 _{e/o})]}
{2  (\epsilon _k^2 + \Delta  ^2 _{e/o})^{1/2}} \simeq \lambda
\bigl(\frac{2\pi k_B T}{ \Delta_{e/o} (0)}\bigr)^{1/2}.
\end{equation}
where $\lambda = |V_{k,k'}|_{av} N(0) {\cal V} \simeq 1/4 $ is the
superconducting coupling constant. The smallest possible value
of the dispersion, $\delta _{e/o} \equiv  E(k = k_F)$, is
slightly different from $\Delta _{e/o}$. In fact, from Eq.(\ref{disp3})
we get
\begin{equation}
\delta _{e/o} = \Delta _{e/o} \Bigl[ 1 \mp \frac{\lambda}{N_{eff}}
\bigl(\frac{2\pi k_B T}{ \Delta_{e/o} (0)}\bigr)^{1/2} \Bigr].
\end{equation}
This means that low temperature behavior of the spectral gap
$\delta _{e/o} $ will be somewhat different from that of $\Delta _{e/o}$
\begin{equation}
\delta_{e}(T) \simeq \Delta_{BCS} (0) \Bigl[ 1 -
\frac{\lambda }{2 \Delta_{o}(0)  N(0) {\cal V}} -
\bigl(4\pi k_B T  N(0) {\cal V} \bigr)
\exp (- 2 \beta \Delta_{e} (0)) \Bigr],
\end{equation}
\begin{equation}
\delta_{o}(T) \simeq \Delta_{BCS}(0) \Bigl[ 1
- \frac{1 - \lambda }{2 \Delta_{o}(0)  N(0) {\cal V}}
- \bigl(\frac{2\pi k_B T}{ \Delta_{o} (0)}\bigr)^{1/2}
\exp (-  \beta \Delta_{o} (0)) \Bigr].
\end{equation}
Clearly, both $\delta_{e}(0) $ and $\delta_{o}(0) $ is smaller than
$ \Delta_{BCS}(0) $
\begin{equation}
\delta_{o}(0) \simeq \Delta_{BCS}(0)
- \frac{\lambda }{2   N(0) {\cal V}  },
\end{equation}
\begin{equation}
\delta_{o}(0) \simeq \Delta_{BCS}(0)
- \frac{1 - \lambda }{2   N(0) {\cal V} }.
\end{equation}
Nevertheless, since $\lambda < 1$, the odd gap is depleted somewhat more
than the even gap. The deviations from the BCS results found in this
section are probably too small to be experimentally
observable in the quantities we investigate later on. Therefore, in the
subsequent calculations we ignore any deviation from the BCS gap equation
and dispersion.

\section{Thermodynamic Properties}
\label{sec:thermo}

Now we can estimate the grand potential itself and the even-odd
grand potential difference in the low temperature limit $ k_B T \ll \Delta $,
by using the standard BCS relations for $E_k$
and $v_k$ in eq.(\ref{f}). In this limit
\begin{eqnarray}
\delta C_{e/o} \simeq \frac{-2}{1 \pm ( 1 + 2 N_{eff} e^{- \beta \Delta})}
\nonumber \\
\times \sum _{k k'} \frac{4 \Delta _k \Delta _{k'}}{E_k E_{k'}}
V_{k k'} e^{- \beta (E_k + 2 E_{k'} )}
\end{eqnarray}
is exponentially small for both even and odd parity. In fact, after neglecting
the parity dependence of $C$, the results obtained within this approximation
do not differ from those obtained from
Eq. (\ref{5}).
For the even-odd grand potential difference, calculated at a fixed
chemical potential $\mu $, we obtain
\begin{equation}
\delta {\Omega}_{e/o} (\mu ) \equiv {\Omega }_o (\mu ) - {\Omega}_e (\mu )
= \Delta - k_B T \ln N_{eff}. \label{fd}
\end{equation}
The experimentally relevant quantity is the free energy difference of
two isolated (canonical) systems with $2N+1$ and $2N$ particles respectively.
This can be related to $\delta {\Omega}_{e/o} $ through the relation
\begin{equation}
e^{- \beta \Omega } = \sum_N e^{- \beta F_N (T)} e^{ \mu \beta N }.
\end{equation}
where $F_N(T)$  is the free energy energy for a canonical ensemble. Using a
Gaussian approximation about the peak of the summand, one obtains
\begin{eqnarray}
\FL
{\Omega }_o (\mu _o ) - {\Omega}_e (\mu _e) \simeq F(2N+1) - \mu_o (2N+1)
+ F(2N) - \mu_e 2N - \frac {1}{2 \beta } \ln \bigg( \frac {\Delta N _o ^2}
{\Delta N _e ^2} \bigg)
\label{omega}
\end{eqnarray}
where $\Delta N _{e/o} ^2$ are the number fluctuations in the grand
ensembles (\ref{5}). The investigation of how the number fluctuations
depend on the number
parity of the ensemble is an important test of the present model,
since one must check that the fluctuations in the
parity restricted grand canonical ensembles do not wash out the free energy
difference one is attempting to calculate.

The number operator written in terms of the Bogoliubov creation and
annihilation operators has the form \cite{schrieffer}
\begin{eqnarray}
 N = \sum _k v^2_k + (u_k^2 - v_k^2) [ \gamma _{k0}^{\dagger } \gamma _{k0} +
                    \gamma _{-k1}^{\dagger } \gamma _{-k1} ]
	+ 2 u_k v_k [ \gamma _{k0}^{\dagger } \gamma _{-k1}^{\dagger } -
                    \gamma _{k0} \gamma _{-k1} ].
\label{nop}
\end{eqnarray}
The last two terms  are off diagonal in quasiparticle number, and therefore
they do not contribute to the average particle number. ( They, however, turn
out to be crucial for the mean square fluctuations
$\Delta N _{e/o}^2$ ). The average quasiparticle number is
\begin{equation}
< N > _{e/o} = \sum _k v _k^2 + (u_k^2 - v_k^2)
\frac {f_+ {\cal Z}_+ \pm f_- {\cal Z}_-}{{\cal Z}_+ \pm {\cal Z}_-}.
\label{nbar}
\end{equation}
Now we can fix the chemical potentials for the even/odd systems in
(\ref{omega}), since they satisfy the above equation
with $ <N>_e = 2N $ and $ <N>_o = 2N+1$, respectively. From (\ref{nbar})
we obtain that $ \mu _o = \mu _e + \delta \mu $, where
$\delta \mu \propto \epsilon _F / 2N$ is of the order of the level spacing
(much smaller than $\Delta $).

When calculating the mean square fluctuations $\Delta N _{e/o}^2 =
< N ^2 >_{e/o} - < N >_{e/o} ^2$ by using the operator expression
in eq. (\ref{nop}, we see that the off-diagonal terms now give a nonzero
contribution, proportional to $ ( \Delta / E_k )^2 $
\begin{eqnarray}
\FL
\Delta N _{e/o}^2 = \sum_k \bigl\{ 2 \bigl(\frac{\epsilon _k}{E_k} \bigr)^2
\bigl[ f_+ ( 1 - f_+ ) - \frac {f_+ ^2 - f_- ^2}{1 \pm {\cal Z}_- /
{\cal Z}_+}\bigr]
\nonumber\\
+ 4 \bigl(\frac{\Delta }{E_k} \bigr)^2
\bigl[ f_+ ^2 + ( 1 - f_+ )^2 - \frac {f_+ ^2 - f_- ^2}{1 \pm {\cal Z}_-
/ {\cal Z}_+}
- \frac {(1 - f_+ )^2 - (1 - f_- )^2}{1 \pm {\cal Z}_- /
{\cal Z}_+}\bigr] \bigr\}.
\end{eqnarray}
At low temperatures, when $ f_{\pm } \rightarrow \pm \exp(- \beta \Delta ) $,
direct calculation shows that:
\begin{equation}
\Delta N _{e/o}^2 = 4 \sum_k (\Delta / E_k )^2  +
{\cal O} (1/N_{eff}) ,
\label{fluct}
\end{equation}
{\it regardless } of the even or odd nature of the
ensemble. This very important result reassures us that thermal and quantum
fluctuations in the particle number do not dominate the difference between
the grand canonical ensembles of even and odd parity in Eq.(\ref{omega}).
Therefore this
modified grand canonical method can be used to calculate even/odd
differences in canonical quantities. One should note that this property
is intrinsic to the superconducting state: at $T = 0$ the fluctuations
are of a purely  quantum nature
$ \Delta N _{e/o} \propto \sqrt{N \Delta / E_F}$,
and decrease as the superconducting gap $\Delta $ is depleted. In contrast,
the same quantity for the normal state is
$ \Delta N _{e/o} \propto \sqrt{N T / T_F} \rightarrow 0$ as
$ T \rightarrow 0$.

Expanding ${\Omega }_e (\mu _e ) \simeq
 {\Omega }_e (\mu _o ) + \delta \mu 2N$ on the left-hand side of
Eq.(\ref{omega}), and using (\ref{fluct}) and (\ref{fd}), we obtain :
\begin{equation}
F(2N+1) - F(2N) - \mu = \Delta - k_B T \ln N_{eff}.
\end{equation}
This expression, previously given in refs. \cite{tuominen,devoret},
has been used to provide a reasonable account of the highest temperature
at which the $2e$ periodicity of the I-V characteristics for the
single-electron-transistor is experimentally observed by Tuominen and
collaborators \cite{tuominen}:
$ T^* = \Delta / k_B \ln N_{eff} \sim 300 $ mK. It also
agrees with the temperature dependence of $ \delta {\cal F}_{e/o} $
measured by Lafarge et al.\cite{devoret}.

The method presented here allows us to calculate other possibly
measurable properties of the superconducting state with restricted
particle number parity. For example, we can calculate the entropy from
the expression
\begin{equation}
S_{e/o} = - k_B Tr \{ \rho _{e/o} ln \rho_{e/o} \},
\label{en}
\end{equation}
where the density matrix is given by
\begin{equation}
\rho_{e/o} = {\cal Z}_{e/o} ^{-1}  \frac{1 \pm (-1)^N}{2}
e^{- \beta ({\cal H} - \mu N)}  .
\end{equation}
After performing the parity restricted trace and using the results we
obtained previously for ${\cal Z}_{e/o}$ ( Eq. (\ref{part})) we get
\begin{eqnarray}
\FL
S_{e/o} = S_{BCS} + \frac{1}{T}\sum E_k \delta f_{e/o} +
k_B \ln \bigg[ \frac{1}{2}
\Big( 1 \pm \prod _{k, \sigma }
\tanh \frac{ \beta E_k}{2} \Big) \bigg] ,
\end{eqnarray}
where $S_{BCS}$ is the entropy of a BCS superconductor
\begin{equation}
S_{BCS} = - k_B \sum _{k, \sigma }
\{ (1 - f_+) \ln (1 - f_+) + f_+ \ln f_+ \} ,
\end{equation}
and $\delta f_{e/o} $ is the even/odd change in the quasiparticle
occupation $f_+ $
\begin{equation}
\delta f_{e/o} = \frac{ f_+ - f_-}{1 \pm {\cal Z}_+/{\cal Z}_-}
= \frac{1}{ \sinh \beta E_k}
\frac{ 1}
{1 \pm \prod _{k \sigma} \coth \frac{\beta E_k}{2}}.
\end{equation}

In the low temperature limit the entropy
for the even-parity system is
\begin{equation}
S^{e} \simeq k_B \beta \Delta N_{eff}^2
e^{ - 2 \beta \Delta},
\end{equation}
whereas for the odd-parity system we have
\begin{equation}
S^{o} \simeq k_B \ln N_{eff} + S_{BCS}.
\end{equation}
Here $S_{BCS} = k_B N_{eff} (\beta \Delta) e^{ - \beta \Delta}$.
The specific heat \cite{devoretm} can be obtained directly from the
above expression via
\begin{equation}
C_v^{e/o} = T \frac{ \partial S_v^{e/o} }{ \partial T}
\end{equation}
More specifically, for the even case
\begin{equation}
C_v^{e}  \simeq k_B (N_{eff} \beta \Delta )^2
e^{ - 2 \beta \Delta}.
\end{equation}
Note the $\exp (- 2 \beta \Delta )$ dependence of the heat
capacity for this case, different from the usual $\exp (- \beta \Delta )$
in the BCS case. This discrepancy is the direct consequence of the number
parity restriction applied to the system. The odd-parity system
has a specific heat that is finite at all temperatures due to the
presence of an unpaired electron:
\begin{equation}
C_v^{o}  \simeq \frac{1}{2} k_B + C_v^{BCS}.
\end{equation}

Finally, we would like to make some technical remarks  regarding
the calculations presented in this paragraph.
The result in Eq. (\ref{nbar}) can be obtained in another way as
well, by using the thermodynamic relation
$ < N >_{e/o} =  - \partial \Omega _{e/0}/ \partial \mu $ keeping
in mind the requirements from minimization, $ \partial \Omega _{e/0}/
\partial E_k = 0  $ and  $ \partial \Omega _{e/0}/ \partial v_k = 0 $ .
In contrast to this, the thermodynamic relationship $ \Delta N _{e/o}^2
= \frac{1}{\beta } \partial < N > / \partial \mu $  does {\it not} give
the correct result for the mean square fluctuations: the crucial off-diagonal
terms do not contribute to $< N >$ and as a result are missed in the expression
for $\Delta N _{e/o}^2 $ derived in this way. This reflects the peculiarity
of the superconducting state.
Yet another possible source of error is encountered if one tries
to calculate the entropy by directly applying the thermodynamic
relation $ S = - (\partial \Omega / \partial T )_V $ instead of using
the expression (\ref{en}). The reason is the following: the temperature
dependence of the function $C = <{\cal H}_I >$ is an artifact of the
mean field approximation. Therefore, when calculating the entropy in this
way, one should {\it not }
take the derivative of $C$. In fact, by evaluating $S_{e/o}$ with the
help of Eq.( \ref{en}), one can explicitly show that within the mean
field approximation all contributions from $C_{e/o}$ drop out. Exactly the
same situation arises in the ordinary BCS calculation.

\section{Transport properties}
\label{sec:trans}

Here we calculate the response of
the parity restricted superconducting state to
various external probes. The acoustic attenuation rate
of ultrasound in a superconductor, within the framework of  the BCS
theory \cite{schrieffer}, is given by the  expression
\begin{eqnarray}
\FL
\alpha_S (T) = \frac{2 \alpha_N (T)}{\hbar \omega} \int _{\Delta } ^{\infty }
dE \frac{E}{\sqrt{E^2 - \Delta ^2}}\frac{E'}{\sqrt{(E')^2 - \Delta ^2}}
\bigg( 1 - \frac{\Delta ^2 }{E E'}\bigg) [ f_+(E) - f_+(E')],
\end{eqnarray}
where $E' = E + \hbar \omega$.
In the parity restricted ensemble the thermal factor is replaced by
\begin{eqnarray}
\FL
f_+(E) - f_+(E') \rightarrow
\frac{[f_+(E) - f_+(E')]{\cal Z}_+ \pm [f_-(E) - f_-(E')]{\cal Z}_-}
{{\cal Z}_+ \pm {\cal Z}_-}.
\end{eqnarray}
As a result, we obtain for the ratio of the rates
in the superconducting and the normal phase
\begin{equation}
\Bigl( \frac{\alpha_S (T) }{\alpha_N (T) } \Bigr)_{e/o}
= 2 f(\Delta ) - \frac{2}{1 \pm
\prod_{k, \sigma} \coth \frac{ \beta E_k}{2}}
\frac{1}{\sinh  \beta \Delta }.
\end{equation}
In the low temperature limit, for the even case  the ratio
drops faster than in the BCS case
\begin{equation}
\Bigl( \frac{\alpha_S (T) }{\alpha_N (T) }\Bigr)_e =
 2 N_{eff}e^{ - 2 \beta \Delta }.
\end{equation}
This should be contrasted with the $\exp ( - \beta \Delta )$ dependence
of the BCS ratio. The odd case also shows a deviation
\begin{equation}
\Bigl( \frac{\alpha_S (T) }{\alpha_N (T) } \Bigr)_0 = 2 e^{ - \beta \Delta }
+ \frac{2}{ N_{eff}}.
\end{equation}
Both terms are of the same order of magnitude at $T^*$, the onset temperature
of even-odd effects. As we will see later on the low temperature response
functions to other external probes, like the NSR relaxation rate and
electromagnetic absorption, will have similar behavior as the ultrasound
attenuation calculated above.

The BCS result for the nuclear spin relaxation rate is \cite{schrieffer}
\begin{equation}
\frac{R_S (T) }{R_N (T) } = 2 \int _{\Delta }^{\infty }
\frac{[E E' + \Delta ^2 ] f_+(E) (1 - f_+(E')}
{[E^2 - \Delta ^2 ]^{1/2}[(E')^2 - \Delta ^2 ]^{1/2} }.
\end{equation}
Within the even/odd ensemble, the thermal factor is modified to
\begin{eqnarray}
\FL
f_+(E)( 1 - f_+(E')) \rightarrow
\frac{f_+(E){\cal Z}_+ \pm f_-(E) {\cal Z}_-}{{\cal Z}_+ \pm {\cal Z}_-} -
\frac{f_+(E)f_+(E'){\cal Z}_+ \pm f_-(E)f_-(E') {\cal Z}_-}
{{\cal Z}_+ \pm {\cal Z}_-}.
\end{eqnarray}
An explicit estimation shows that only the first term will contributes
significantly, and the second can be neglected.
Denoting with $F_{e/o}(E)$ the low temperature, $E \rightarrow E'$ limit
of the new parity dependent thermal factor, we obtain
\begin{eqnarray}
F_e(E) = N_{eff}e^{ - \beta (E + \Delta) } ,\\
F_o(E) = e^{ - \beta E}
+ \frac{e^{ - \beta (E - \Delta) }}{ N_{eff}}.
\end{eqnarray}
Using this function, the ratio of the relaxation rates can be written as
\begin{equation}
\Bigl( \frac{R_S (T) }{R_N (T) } \Bigr)_{e/o}= 2 F _{e/o}(\Delta )
\int _{\Delta }^{\infty }
\frac{[E^2  + \Delta ^2 ] e^{ - \beta (E - \Delta)}}
{[E^2 - \Delta ^2 ] },
\end{equation}
where the integral now is only a weak function of temperature. All the
important temperature and parity dependence is embodied in
$F_{e/o} (\Delta )$.  Also note that $F_{e/o} (\Delta )$ gives the low
temperature behavior of the acoustic attenuation as well, since
$ ( \alpha _S / \alpha _N )_{e/o} \simeq 2 F_{e/o} (\Delta )$.

Finally, let us discuss the infrared absorption.
At very low temperatures there are only a few thermally excited
quasiparticles available, and therefore substantial contribution
to the absorption rate is given by the pair breaking process
in which an incoming photon breaks a Cooper pair and creates
two quasiparticles. In the case of the odd-parity system, however,
there is always an unpaired quasiparticle present, which will lead to
nonzero subgap absorption. Let us therefore consider the case when
$\hbar \omega < 2\Delta $, so that only single particle
processes are allowed. We examine
the real part of the conductivity - proportional to the
rate of absorption - which has the following BCS form
\cite{schrieffer,parks}
\begin{equation}
\frac{\sigma _{1S}(T)}{\sigma _{1N}(T) } = 2 \Bigl[ f_+ (\Delta )
+ \Delta ^2 \int _{\Delta }^{\infty }
\frac{[f_+(E) (1 - f_+(E)}
{E^2 - \Delta ^2}\Bigr], \ {\rm for } \  \hbar \omega < 2 \Delta.
\end{equation}
A calculation similar to that presented above for the nuclear spin
relaxation gives
\begin{equation}
\frac{\sigma _{1S}(T)}{\sigma _{1N}(T) } = 2 F_{e/o} (\Delta )
\Bigl[ 1 + \Delta ^2 \int _{\Delta }^{\infty }
\frac{e^{ - \beta (E - \Delta)}}
{E^2 - \Delta ^2}\Bigr], \ {\rm for } \ \hbar \omega < 2 \Delta.
\end{equation}

The common feature of all these results obtained for the response
of the parity-restricted superconductor to various probes is, that
for low temperatures and energy transfers the $\exp ( -  \beta \Delta )$
dependence - characteristic of a  BCS superconductor - is replaced
by a more rapidly decaying $\exp ( - 2 \beta \Delta )$ behavior for the
{\it even} case, and the appearance of a very small ($ {\cal O} (1/ N_{eff})$),
but nonzero background due to a {\it  single}, unpaired quasiparticle,
that can scatter, absorb, spin relax etc.
Pictorially, the $\exp ( - 2 \beta \Delta )$ dependence comes from the
exclusion of the parity-violating grand-canonical processes  that involve
a single particle exchange between the system and a particle bath
(excitation energy $ \Delta $), as well as  from the important
contribution of the momentum-conserving excitation process of a pair
(excitation energy $2 \Delta $) , in contrast to pair breaking with
no momentum conservation (excitation energy $ \Delta $ per particle).

\section{Conclusions}
\label{sec:concl}

In conclusion, we have presented a theoretical method to investigate
number parity effects in the superconducting state by using a grand
canonical ensemble in which the parity of the system is
even or odd. This corresponds to a permeable wall between the superconductor
and the particle bath that allows particle exchange only in pairs. We obtained
the partition function which in turn yielded the grand potential, entropy and
specific heat of the system. While the variational equations from the
minimization of the grand potential gave essentially BCS results, the free
energy difference between even and odd parity states obtained is in accordance
with the experimental and theoretical findings of Tuominen et al.
and the measurements of Lafarge et al. \cite{tuominen,devoret}. The specific
heat shows a more rapid $ \exp ( - 2\beta \Delta )$ decay at low temperatures;
the same quantity in the odd-parity system has a finite zero temperature
limit $k_B /2 $, in contrast to the usual BCS result.

A simple and systematic low temperature approximation was made possible
by the fact that for the temperature range and size of the
islands investigated so far\cite{tuominen,devoret} $  N_{eff} \gg 1 $ and
therefore an expansion was possible in powers of $1/N_{eff}$. In principle,
this expansion breaks down at extremely low temperatures, since $1/N_{eff}
\propto T^{-1/2} $. In order to observe such effects in the
temperature range presently investigated ($ T \ge 50 \ mK$), the volume
of the island must be four orders of magnitude less,
about $10^{-19} {\rm cm}^3$, which might be very difficult to achieve
experimentally.

We have not addressed the problem of the magnetic field dependence
of the even-odd free energy difference. The measurements of Tuominen et al.
\cite{tuominen2} and Esteve et al. seem to suggest that the experimental
results can be explained if one replaces in eq. (\ref{fd}) the gap in zero
field $\Delta $ with the gap edge $ \Delta (H) $ of the field dependent
density of states provided by the Abrikosov - Gorkov theory \cite{abr}. We
believe that this replacement, although possibly correct, is not
justified well enough. The
present method cannot be easily generalized to the Green's function
technique, due to the lack of a Wick's theorem: for the parity-conserving
average defined in eq. (\ref{av}),
$\langle \hat n_k \hat n_{k'} \rangle \ne \langle \hat n_k \rangle
\langle \hat n_{k'} \rangle $ for a system of noninteracting
fermions, where $ \hat n_k $ is the number operator
(cf. Eq.(\ref{rho})).
There is, however, a Wick's theorem for the two channels, with projection
operator $1/2$ and $ \pm (-1)^N /2$, respectively, which add up to the
parity-restricted average. Because of this, the normal-state properties
and the Cooper instability need further investigation, to be performed
in the future.

We thank M. Devoret, D. Esteve, K. A. Matveev, and A. Zawadowski
for helpful comments and correspondence.
One of us (A. S.) gratefully acknowledges a grant from the Danish Research
Academy. This work was supported in part by the MRL Program of the NSF under
award No. DMR-9121654.


\end{document}